\documentclass[showpacs,twocolumn,pre]{revtex4}
\usepackage{psfrag,epsfig,amsfonts,amssymb,amsmath,graphicx,slashbox}
\usepackage{dcolumn}

\begin{document}

\title{The Brownian gyrator: a minimal heat engine on the nano-scale}

\author{Roger Filliger}
\affiliation{Berne University of Applied Sciences, Engineering and
Information Technology, 2501 Biel/Bienne, Switzerland}

\author{Peter Reimann}
\affiliation{Universit\"at Bielefeld, Fakult\"at f\"ur Physik, 33615 Bielefeld, Germany}

\begin{abstract}
A Brownian particle moving in the vicinity of a generic
potential minimum under the influence of
dissipation and thermal noise from two
different heat baths is shown to act
as a minimal heat engine, generating a systematic torque
onto the physical object at the origin of the potential
and an opposite torque onto the medium generating the
dissipation.
\end{abstract}

\pacs{05.40.-a, 05.60.-k, 05.70.Ln}

\maketitle {\em Introduction and summary:} The theory of heat
engines is one of the main roots of modern thermodynamics and
statistical physics. Recently, there has been a considerable
renewed interest in the long standing problem of determining the
fundamental efficiency limit of a heat engine at maximum power
\cite{ref1}. Another exciting new perspective is the conceptual
design of a cooling device on the nano- or even single-molecule
scale by inverting a heat engine which is powered by Brownian
motion \cite{ref2}. The original and still paradigmatic setup of
a heat engine consists of two heat baths in contact with a
cyclically working ``engine'', generating work in the form of a
torque. Here, we put forward the smallest and most primitive such
engine one may think of: a single particle, gyrating around a
generic potential energy minimum under the influence of friction
and thermal noise forces from two simultaneously acting heat
baths. Three typical examples are indicated in Fig. 1. While the
particle itself is way too small to store or disburse any notable
amount of angular momentum, it acts as a kind of catalyst. The
particle in fact generates, with the help of the disequilibrium
between the two baths, a systematic average torque onto the
physical object at the origin of the potential and an opposite
torque of the same magnitude by way of the dissipation mechanism
onto one or both heat baths.

Such Brownian gyrators are minimal heat engines in so far as they
are acting essentially like their macroscopic counterparts, but
the ``engine'' itself needs not be anything more than a
structureless particle performing Brownian motion. Due to this
simplicity they are in principle readily realizable by
nano-technological or even single-molecule techniques. Potential
applications are manifold and obvious: wrapping and unwrapping of
DNA and other polymers \cite{mob06}, driving wheels or screws of
nano-devices \cite{ref2} and synthetic molecular motors
\cite{kay07}, stirring and mixing in micro- and nano-fluidic
devices \cite{squ05}, to name but a few. Though each single
engine is ``weak'', when acting in parallel, a case naturally
arising in the context of colloidal particles \cite{magnet,light}
or magnetic fluxons \cite{fluxon}, the result may even be a
macroscopic torque \cite{eng03}.

{\em Model:}
Focusing on the simplest case, we consider
the motion of a point particle with two
spatial degrees of freedom ${\bf x}:=(x_1,x_2)$
in a static potential $U({\bf x})$.
We assume that the potential has a minimum at
the origin ${\bf x}={\bf 0}$ and that large
excursions are sufficiently rare to admit
a parabolic approximation of the form
\begin{eqnarray}
U({\bf x})= \sum_{i=1}^2 \frac{u_i}{2} y_i^2 \ , \ \
y_i:=\sum_{k=1}^2 O_{ik}(\alpha) x_k \ ,
\label{1}
\end{eqnarray}
where $O(\alpha)$ is a $2 \times 2$ orthogonal
matrix with elements $O_{11}=O_{22}=\cos\alpha$,
$O_{12}=-O_{21}=\sin\alpha$,
describing a rotation in the plane by an
angle $\alpha$.
The transformed coordinates $y_i$
thus correspond to the ``principal axes'' of the
parabolic potential in (\ref{1}) and
$u_1, \, u_2 > 0$ are the corresponding
``principal curvatures''.
Clearly, Eq. (\ref{1}) represents the generic
form of a potential minimum in 2 dimensions.
Furthermore, one has $u_1\not = u_2$ in the
generic case, i.e. unless the system is
rotationally symmetric about the origin.

\begin{figure}
\epsfxsize 0.65 \columnwidth
\epsfbox{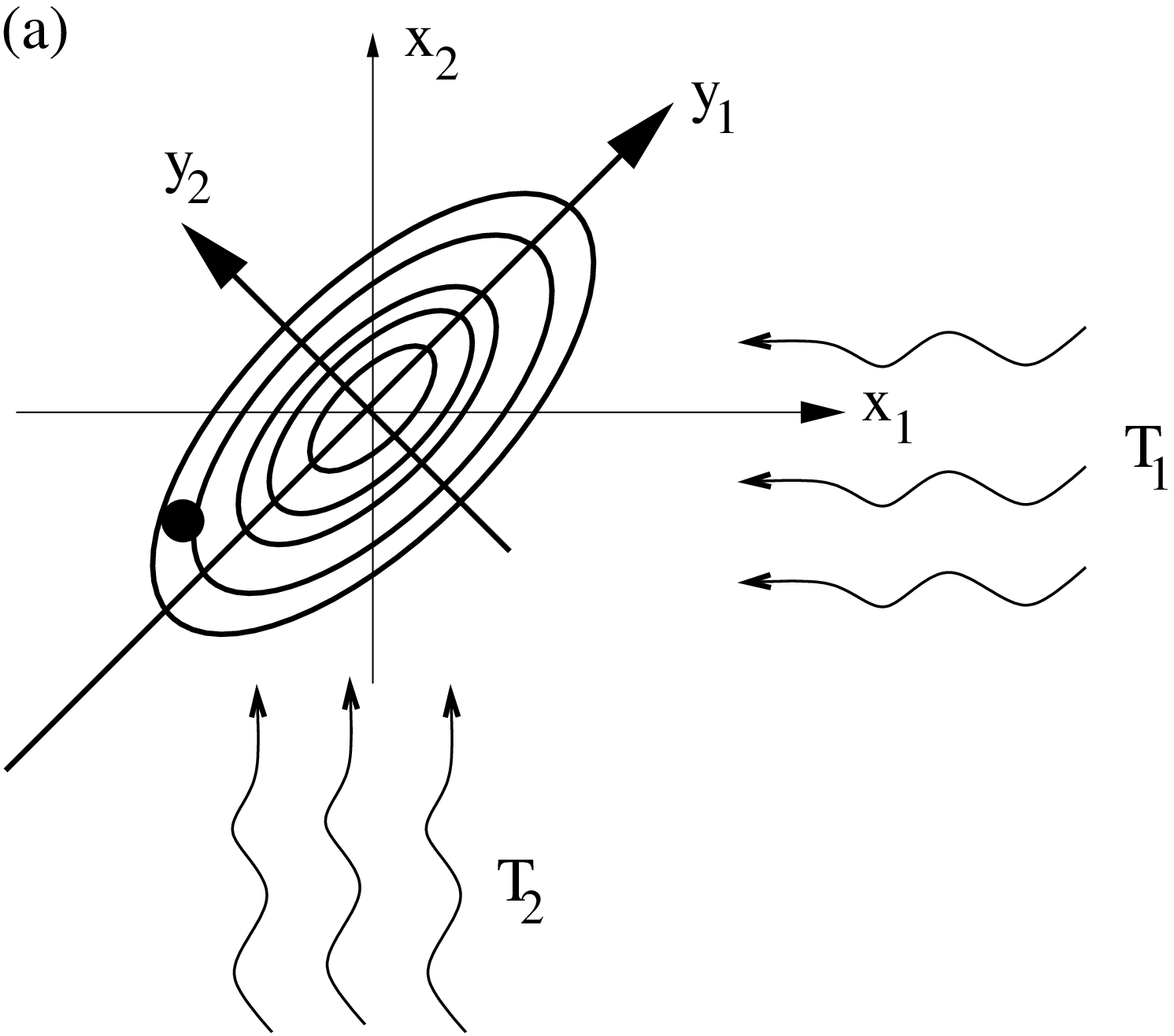}
\\[0.5cm]
\epsfxsize 0.65 \columnwidth
\epsfbox{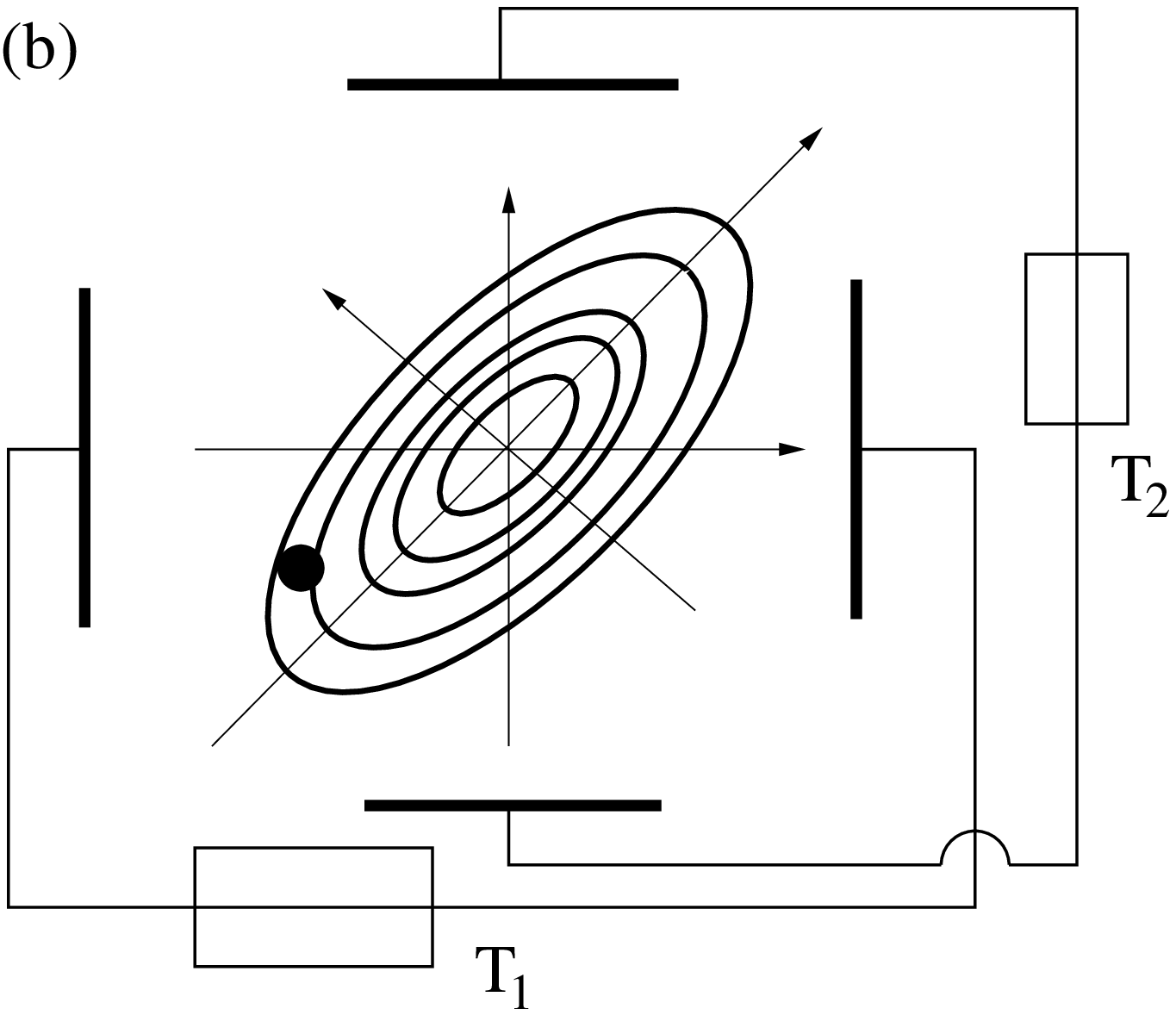}
\\[0.5cm]
\epsfxsize 0.65 \columnwidth
\epsfbox{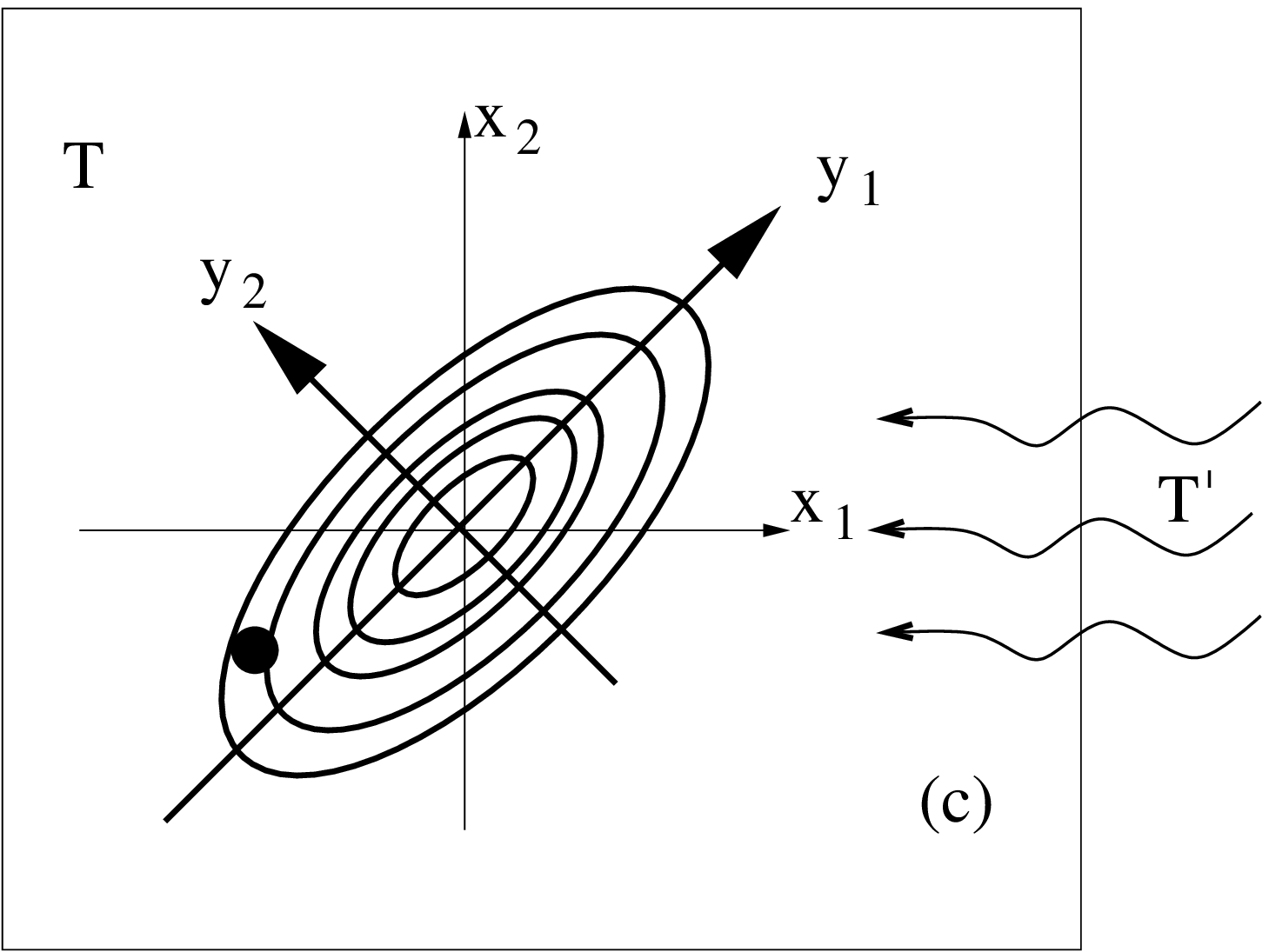}
\caption{
Schematic sketch of various types of
Brownian gyrators acting as heat engines.
The black dot represents the Brownian particle,
moving in the $x_1$-$x_2$-plane.
The contour lines indicate a typical parabolic
potential (\ref{1}) with principal axes
$y_1$ and $y_2$.
(a)-(c) illustrate different realizations of
the two heat baths.
(a): Both heat baths act on the charged
particle by way of black body radiation
at different temperatures $T_1$ and $T_2$,
irradiating along the $x_1$- and $x_2$-axes,
respectively.
The dissipation mechanism is provided by
radiation damping into the vacuum.
(b): ``Electrical heat baths'', realized by
two resistors at different temperatures
$T_1$ and $T_2$, coupled to the charged
particle by means of two plate condensers.
Each of them transfers the random voltage fluctuations
of one resistor to the particle along a preferential
direction and gives rise to dissipation via the resistor
when the particle moves and hence induces a current
in the electrical circuit.
Replacing the condenser plates by Helmholtz coils
gives rise to ``magnetic baths'' interacting with
e.g. a paramagnetic particle \cite{eng03}.
Replacing the condenser plates by piezo elements
gives rise to ``acusto-mechanical'' baths \cite{cou07}.
(c): Only one heat bath (with temperature $T'$)
is of the anisotropic type as
in (a) and (b). The second heat bath (with temperature $T$)
consists of the usual fluid environment of the
Brownian particle.
}
\label{fig1}
\end{figure}

To complete our heat engine, we need two
heat baths which act onto the particle without
resulting in a total equilibrium system.
In the most common case, the heat engine
is alternatingly brought into contact with
two baths at two different temperatures.
Since this requires a quite complicated
machinery in practice, here we rather
focus on the case that the particle is
permanently in contact with both baths.
In the following, we first discuss in detail
the conceptually simplest theoretical model
and only afterwards turn to the experimental
realizations and more general
system classes.

In the simplest case, at least
one of the two heat baths interacts with the
particle along a preferential direction,
which can be identified with the $x_1$-axis
without loss of generality. Generically,
this direction is not related in any particular way
to the ``principal axes'' of the potential in (\ref{1}),
and hence their relative angle $\alpha$ is not bound
to take any particular value a priori.

Whereas one heat bath thus solely influences
the particle motion along the $x_1$-axis,
the second one may either be of isotropic
character or acting only on $x_2$.
For the sake of simplicity, we first focus
on the latter case, see Fig. 1(a,b).
Modeling the thermal bath effects
as usual \cite{rat,ris} by Gaussian white noise
and a concomitant dissipation proportional to the velocity,
and neglecting inertia effects, we arrive at
the following overdamped Langevin equations
for the particle dynamics in the plane:
\begin{equation}
\eta_i \dot x_i(t) = -\frac{\partial U({\bf x}(t)) }{\partial x_i}
+ \sqrt{2\eta_i k_BT_i}\, \xi_i(t)\ , \ i=1,2 \ .
\label{2}
\end{equation}
Here, $k_B$ is Boltzmann's constant, $T_i$ is the temperature of the $i$-th
bath, $\xi_i(t)$ are independent, $\delta$-correlated Gaussian
noises, and the coupling strength between particle and bath $i$
is quantified by the friction coefficient $\eta_i$ \cite{rat,ris}.

{\em Solution:}
We first discuss in some more detail the forces and
torques connected with the dynamics (\ref{2}).
Denoting by ${\bf e}_i$ the unit vector along the
$i$-th coordinate axis, the three relevant forces are:
the dissipation ${\bf f}_\eta(t) := -\sum_{i=1}^2{\bf e}_i \eta_i \dot x_i(t)$,
the potential force ${\bf f}_U(t):=-\nabla U({\bf x}(t))$,
and the fluctuation force
${\bf f}_\xi(t):= \sum_{i=1}^2{\bf e}_i \sqrt{2\eta_i k_BT_i}\, \xi_i(t)$.
Hence, (\ref{2}) is tantamount to the force balance
${\bf f}_\eta(t)+{\bf f}_U(t)+{\bf f}_\xi(t)={\bf 0}$.
On the average over many realizations of the noise
we thus obtain $\langle {\bf f}_\xi(t)\rangle = {\bf 0}$ and hence
$\langle {\bf f}_U(t)\rangle = - \langle {\bf f}_\eta(t)\rangle$,
corresponding to the following elementary physics:
Since the particle momentum is by definition considered as
negligible in the overdamped limit, the force
exerted by the potential $U$ is compensated on
the average by the friction forces.
A similar consideration applies to the three torques
of the form ${\bf f}\times {\bf x}$: The thermal
fluctuations do not give rise to any systematic torque,
$\langle{\bf f}_\xi(t)\times {\bf x}(t)\rangle={\bf 0}$, and hence
the torque of modulus $M(t)$ and direction
${\bf e}_3:={\bf e}_1\times{\bf e}_2$
which the particle exerts on the potential $U$
(or the physical object at the origin of
that potential) is on the average exactly equal to the
opposite torque $-M(t){\bf e}_3$ which the particle
exerts via the friction forces on the thermal environment
(or the physical objects containing the baths):
\begin{equation}
\langle{\bf f}_U(t)\times {\bf x}(t)\rangle
= - \langle{\bf f}_\eta(t)\times {\bf x}(t)\rangle
=M(t)\,{\bf e}_3 \ .
\label{3}
\end{equation}

Next we turn to the Fokker-Planck-equation
\cite{rat,ris} equivalent to (\ref{2}),
\begin{equation}
\frac{\partial P({\bf x}, t)}{\partial t}
=-\sum_{i=1}^2\frac{\partial J_i({\bf x}(t),t)}{\partial x_i} \ ,
\label{4}
\end{equation}
where $P({\bf x}, t)$ is the probability density to
find the particle at position ${\bf x}$ at time $t$ and
${\bf J}=(J_1,J_2)$ the corresponding probability
current density with components
\begin{equation}
J_i({\bf x}, t)=-\left[\frac{1}{\eta_i}\frac{\partial U({\bf x})}{\partial x_i}
+\frac{k_BT_i}{\eta_i}\frac{\partial}{\partial x_i}\right]\, P({\bf x},t) \ .
\label{5}
\end{equation}
The Fokker-Planck-equation (\ref{4},\ref{5}) is
complemented by natural boundary conditions
$P({\bf x},t)\to 0$ and $J_i({\bf x},t)\to 0$ for $x_i\to\pm\infty$.
Given $P({\bf x}, t)$, the torque modulus $M(t)$ from (\ref{3})
readily follows according to
\begin{equation}
M(t)=\int P({\bf x}, t)
\left(
x_1\frac{\partial U({\bf x})}{\partial x_2} -x_2\frac{\partial U({\bf x})}{\partial x_1}
\right)
\,  dx_1\, dx_2
\label{6}
\end{equation}

After initial transients have died out, the system approaches a
unique, steady probability density as $t\to\infty$
\cite{ris,rat,villani}.
For the parabolic potential (\ref{1}), this unique steady state
solution of the Fokker-Planck-equation (\ref{4},\ref{5}) can be
obtained in closed analytical form.
Since the calculations are straightforward but rather tedious
and the expressions quite bulky and not very
illuminating, they are not explicitly given here.
Rather we immediately present the resulting
torque (\ref{6}) in the steady state, reading
\begin{equation}
M = \frac{k_B(T_1-T_2)\, (u_1-u_2)\, \sin 2\alpha}
{u_1+u_2 - \frac{\eta_1-\eta_2}{\eta_1+\eta_2}\, (u_1-u_2) \cos 2\alpha} \ .
\label{10}
\end{equation}
Note that for symmetry reasons, any value of the potential force
${\bf f}_U(t)$ occurs with the same probability
as its inverse in the steady state.
On the average, we thus have $\langle{\bf f}_U(t)\rangle={\bf 0}$,
implying as usual that the resulting torque (\ref{10}) remains
unchanged for any other choice of the reference rotation
axis in (\ref{3}).

{\em Discussion:}
The closed, general expression (\ref{10}) for the average
torque of a Brownian gyrator
in the steady state is the main result of our paper.
For $T_1=T_2$ we are dealing with an equilibrium system
in (\ref{2}) and hence the average torque must vanish
due to the second law of thermodynamics.
If $u_1=u_2$ the potential (\ref{1}) is rotationally symmetric
and hence there cannot be any preferential direction of rotation
and the torque must vanish.
If $\sin 2\alpha =0$ then the principal axes of
the parabolic potential agree with the directions along
which the two heat baths in (\ref{2}) are acting, hence
a net torque is again ruled out by symmetry.
In any other case a finite torque (\ref{10}) results
(the denominator is always positive
since $u_i,\,\eta_i>0$).
A particularly simple behavior arises for equal
coupling strengths $\eta_1=\eta_2$. In this case, the
maximal torque is reached at $\alpha=\pi/2+n\pi$.
In general, the optimal angles will be slightly different.
Typically, the maximal torque is roughly
given by the difference between the thermal
energies associated with the two baths,
$k_B T_1- k_B T_2$, indicating that the
Brownian gyrator transforms its random motion
into a systematic torque quite effectively.
We, however, remark that speaking about efficiencies
in the usual sense is not possible as long as one
does not know the resulting rotation speed of the
thermal baths relatively to the ``carrier''
of the potential $U$, which is is beyond the
scope of the general model (\ref{2}).

The basic physical origin of the preferential rotation
of the Brownian particle in one direction can be most
easily understood in the limit that one temperature
vanishes, say $T_2=0$ in (\ref{2}). Now, let us
assume the particle has reached (in whatever way) a
position on the $x_1$-axis, i.e. $x_2(t)=0$. In the generic
case that the $x_1$-axis does not coincide with a
principal axis of the parabolic potential in (\ref{1}),
there will be a non-vanishing deterministic force
$-\partial U(x_1(t),0)/\partial x_2$
proportional to $x_1(t)$ acting on the particle along the
$x_2$-direction. Since the noise along this
direction vanishes in (\ref{2}), we
can conclude that the particle is able to cross
the $x_1$-axis only in one direction for all
positive $x_1(t)$-values and in the opposite
direction for all negative $x_1(t)$-values.
In other words, the particle rotates around the
origin in a preferential direction.
It is plausible that qualitatively an analogous
behavior is expected also for finite $T_2$
(different form $T_1$), though the details
will be more complicated.

Some of the above basic physical principles
governing the behavior of a Brownian gyrator
are similar to those governing so-called
thermal ratchets and Brownian motors \cite{rat}.
Yet, to the best of our knowledge, neither of
those ratchet systems are immediately comparable
to the setup treated here, nor are the general
concepts in the context of ratchet effects
of any help to gain easier or deeper
insight in the present case.

{\em Experimental realizations:}
The setup from Fig. 1(a) involving anisotropic black body
radiation is of considerable conceptual interest.
The main practical problems are the weak coupling
of a charged particle to the electromagnetic irradiation
and the vacuum, and that the model (\ref{2}) itself
is a very crude description of the real system.
Yet, the basic concept may well be of relevance
for various dynamical processes in astrophysics,
space physics, and laboratory plasmas,
see \cite{per07} and references therein.
More easy to realize in the lab is the setup
from Fig. 1(b), whose heat baths
consist of simple resistors at different temperatures.
If the so generated thermal fluctuations are
still too weak, an electronic amplification is
straightforward \cite{cou07}.
The shortcoming of this setup is that a conversion
of the torque into a relative rotation between
potential and baths is not desirable, since that
would change the angle $\alpha$ in (\ref{1}).

Most attractive from the experimental viewpoint seems to be the
setup from Fig. 1(c): One heat bath is given -- as usual in the
context of Brownian motion -- by a fluid environment of the
particle without any kind of anisotropy. For later use, we denote
its temperature by $T$ and the Stokes friction coefficient by
$\eta$. They both appear as usual \cite{rat,ris} in both
components of the 2-dimensional dynamics (\ref{2}). Only the
second bath continues to emit its thermal fluctuations along a
preferred direction onto the particle, say along the
$x_1$-dynamics. Paradigmatic examples are \cite{eic05} the
(almost) black body irradiation from the sun or some analogous
artificial device in the lab or any kind of anisotropic (almost)
white noise in the original mechano-acoustical sense, emitted
e.g. by a loudspeaker in the experimental work \cite{cou07}.
Denoting its temperature and dissipation
coefficient by $T'$ and $\eta'$, respectively, the resulting
2-dimensional dynamics is again of the overdamped Langevin type
and can be readily brought into the form (\ref{2}) by means of
the identifications $\eta_2=\eta$, $T_2=T$, $\eta_1=\eta+\eta'$,
$T_2=(\eta T+\eta' T')/(\eta+\eta')$. The resulting torque
(\ref{10}) thus takes the form
\begin{equation}
M = \frac{\frac{\eta'}{\eta+\eta'}\, k_B(T'-T) \,  (u_1-u_2) \, \sin 2\alpha}
{u_1+u_2 - \frac{\eta'}{2\eta+\eta'}\, (u_1-u_2) \cos 2\alpha} \ .
\label{11}
\end{equation}
Typically, the coupling strength $\eta$ to the
fluid will be very much larger than the
dissipation coefficient $\eta'$ due to the
anisotropic second heat bath.
In order that the relevant strength
\begin{equation}
g:=k_BT'\eta'
\label{12}
\end{equation}
of the concomitant anisotropic fluctuations is
non-negligible in spite of the small $\eta'$-value,
the temperature $T'$ must be very much larger than $T$.
Then the resulting torque from (\ref{11}) simplifies
in very good approximation to
\begin{equation}
M = \frac{g}{\eta}\, \frac{u_1-u_2}{u_1+u_2} \, \sin 2\alpha \ .
\label{13}
\end{equation}
Here, any non-zero value of $M$ indicates that the
particle is transferring torque from the
potential $U$ to the dissipative mechanism,
while its own angular momentum always remains negligible.
The effect of that pair of opposite torques will be to
generate rotations of the physical object carrying the
potential and the fluid around the particle
into opposite directions, just in the way any
``engine'' is commonly supposed to operate.

The experimental realization of the potential $U$
is possible in many straightforward ways,
e.g. by means of electro-  or magnetostatic forces
\cite{magnet}, dielectrophoretic effects
(including light forces as exploited in optical traps)
\cite{light}, pinning centers of fluxons \cite{fluxon}, etc.

{\em Outlook:}
We close with some generalizations and perspectives.
First of all, basically the same behavior is expected
when working in 3 rather than 2 dimensions.

More challenging is to better understand
the role of the anisotropy of at least one of
the heat baths with the main goal of possibly
abandon this condition.
The first purpose of this anisotropy is of
mainly technical character.
Namely, within the usual modeling of thermal fluctuations
as Gaussian white noise and the concomitant
dissipation proportional to the instantaneous velocity,
see (\ref{2}), two isotropic baths are effectively
equivalent to one single equilibrium environment with
properly adapted effective friction and temperature.
Hence the anisotropy is indispensable in order to
obtain a non-equilibrium system within this standard
modeling of the thermal baths.
However, there are in principle many possibilities --
some of mainly conceptual interest,
others of practical relevance but mathematically more
difficult to handle -- to model a thermal bath
in a different way, e.g. by means of correlated
Gaussian noise and concomitant memory friction
\cite{rat}.
Two of these baths at different temperatures
can no longer be mathematically transformed to
one single equilibrium bath and hence in this
regard the anisotropy is no longer needed.
A second more fundamental role of the anisotropy
is to break the symmetry between gyrating
clock- and counterclock-wise.
Clearly, breaking this symmetry is an indispensable
pre-requisite of making the Brownian gyrator
work. Hence, in the presence of two isotropic heat
baths, this symmetry must be broken in some other
way.
The most straightforward possibility is
via the potential $U$, e.g. by keeping terms
up to cubic order in the expansion (\ref{1})
of the potential about its minimum.

An interesting extension of the present work will be
to explore the collective phenomena due to
many interacting Brownian gyrators, in particular
the similarities and differences as compared to
collective effects of rotating molecular motors in
membranes \cite{len03} and rotating magnetic discs
confined to a two-dimensional interface
\cite{gry00}.
\begin{center}
\vspace{-6mm}
---------------------------
\vspace{-5mm}
\end{center}
This work was supported by
the Alexander von Humboldt Stiftung
and the Deutsche Forschungsgemeinschaft under
SFB 613 and RE 1344/3-1.



\begin{thebibliography}{10}

\bibitem{ref1}
C. Van den Broeck, Phys. Rev. Lett. {\bf 95}, 190602 (2005);
B. Jimenez de Cisneros and A. C. Hernandez, {\em ibid.} {\bf 98},
130602 (2007)

\bibitem{ref2}
C. Van den Broeck and R. Kawai, Phys. Rev. Lett. {\bf 96}, 210601 (2006);
News item in Nature Physics {\bf 2}, 433 (2006)

\bibitem{mob06}
W. M\"obius, R. A. Neher, and U. Gerland, Phys. Rev. Lett. {\bf 97}, 208102 (2006)

\bibitem{kay07}
E. R. Kay, D. A. Leigh, and F. Zerbetto, Angew. Chem. Int. Ed. {\bf 46}, 72 (2007)

\bibitem{squ05}
T. M. Squires and S. R. Quake, Rev. Mod. Phys. {\bf 77}, 977 (2005)

\bibitem{magnet}
P. Tierno, T. H. Johansen, and T. M. Fischer,
Phys. Rev. Lett. {\bf 99}, 038303 (2007)

\bibitem{light}
P. T. Korda, M. B. Taylor, and D. G. Grier,
Phys. Rev. Lett. {\bf 89}, 128301 (2002);
K. Mangold, P. Leiderer, and C. Bechinger,
{\em ibid.} {\bf 90}, 158302 (2003)

\bibitem{fluxon}
C. Reichhardt and F. Nori, Phys. Rev. Lett. {\bf 82}, 414 (1999)

\bibitem{eng03}
A. Engel et al., Phys. Rev. Lett {\bf 91}, 060602 (2003);
Phys. Rev. E {\bf 70}, 051107 (2004).

\bibitem{ris}
H. Risken, The Fokker-Planck Equation (Springer, Berlin) 1984.

\bibitem{rat}
P. Reimann, Phys. Rep. {\bf 361}, 57 (2002)

\bibitem{villani}
L. Desvillettes and C. Villani, Comm. Pure Appl.
Math. {\bf 54}, 1 (2001).

\bibitem{per07}
S. Perri et al., EPL {\bf 78}, 40003 (2007).

\bibitem{cou07}
C. Coupier, M. Saint Jean, and C. Guthmann,
EPL {\bf 77}, 60001 (2007)

\bibitem{eic05}
R. Eichhorn and P. Reimann, Europhys. Lett. {\bf 69}, 517 (2005)

\bibitem{len03}
P. Lenz et al., Phys. Rev. Lett. {\bf 91}, 108104 (2003)

\bibitem{gry00}
B. A. Grzybowski, H. A Stone, and G. M. Whitesides,
Nature (London) {\bf 405}, 1033 (2000)

\end{thebibliography}
\end{document}